\documentclass{article}
\usepackage{breckenr2005,graphicx}
\frompage{000} \topage{000}                                              
\def\rightmark{Entropy and Phase Space Density at RHIC}
\def\leftmark{S. Pratt}
 
\title{Correlation Functions: Getting into Shape} 

\authors{
{Scott Pratt}\\[2.812mm]
{\normalsize
\hspace*{-8pt} Department of Physics and Astronomy, \\ 
\hspace*{-8pt} Michigan State University, East Lansing, MI~~48824-1321, USA 
}
}

\abstract{The ability to measure characteristics of source shapes using
  non-identical particle correlations is discussed. Both strong-interaction
  induced and Coulomb induced correlations are shown to provide sensitivity to
  source shapes. By decomposing correlation functions with spherical or
  Cartesian harmonics, details of the shapes can be especially well isolated.}

\keyword{QGP,correlations} 
\PACS{25.75.-q}
 
\begin{document}

\maketitle
\setcounter{page}{1}

The space-time development of the expanding fireball in a heavy-ion collision
at RHIC is driven by the pressure and viscosity of the matter during the novel
stage of the reaction. Insight into these fundamental questions can be gained
by understanding the spatio-temporal aspects of the collision. Correlation
functions \cite{heinzjacak,starhbt,phenixhbt,phoboshbt} provide a direct link
to the space-time structure of the source function, ${\cal
  S}({\bf P},{\bf r})$,
\begin{equation}
\label{eq:master}
R({\bf P},{\bf q})=\int d^3r' 
\left[\left|\phi({\bf q}',{\bf r}')\right|^2-1\right] 
S({\bf P},{\bf r}').
\end{equation}
Here, $R({\bf P},{\bf q})+1$ is the ratio of two-particle cross-sections to
the corresponding quantity for mixed-events, and is zero for uncorrelated
emission. The source function $S({\bf P},{\bf r}')$ measures the
probability that two particles of the same velocity whose total momentum is
${\bf P}$ are separated by ${\bf r}$. The primes denote that the coordinates
are measured in the rest frame of the pair. By measuring $R({\bf P},{\bf q})$
as a function of the relative momentum ${\bf q}'=({\bf p}'_a-{\bf p}'_b)/2$,
one can exploit the structure of the relative wave function $\phi({\bf q}',{\bf
  r}')$ to extract information about $S({\bf P},{\bf r})$.

Understanding the source function in Eq. (\ref{eq:master}) assists in the
determination of collective flow and lifetimes which directly address questions
concerning the equation of state. In this talk the phenomenology that connects
the source function to these important issues will be ignored. Instead, this
talk focuses on the ability to determine size and shape information about
$S({\bf r}')$ from $R({\bf q}')$. Thus, from this point on the total
momentum label and the primes are suppressed. This problem can be considered as
an inversion problem. Both $R({\bf q})$ and $S({\bf r})$ can be
considered as vectors and $|\phi({\bf q},{\bf r})|^2$ can be considered as a
matrix. Simply stated, Given $R({\bf q})$ and $|\phi({\bf q},{\bf r})|^2$, what
can one determine about $S({\bf r})$?  For non-interacting identical
particles, $|\phi|^2=1+\cos 2{\bf q}\cdot {\bf r}$, and one can Fourier
transform $R({\bf q})$ to find $S({\bf r})$. Our goal is to investigate the
analogous deconvolution for wave functions whose forms are driven by Coulomb
and strong interactions.

A straight-forward case to consider is interactions through classical Coulomb
correlations. In the classical limit,
\begin{eqnarray}
\label{eq:classcoul}
|\phi(q,r,\cos\theta_{qr})|^2&\rightarrow& \frac{d^3q_0}{d^3q}\\
\nonumber
&=&\frac{1+\cos\theta_{qr}-x}
{\sqrt{(1+\cos\theta_{qr}-x)^2-x^2}}
\Theta(1+\cos\theta_{qr}-2x),
\end{eqnarray}
where ${\bf q_0}$ is the relative momentum when the particles were separated by
${\bf r}$, $\theta_{qr}$ is the angle between ${\bf q}$ and ${\bf r}$, and
$x=2\mu e^2/q^2r$. When integrating over angles, $|\phi|^2=\sqrt{1-x}$, and the
correlation function approaches unity at large $q$ as $1-\langle 1/r\rangle \mu
e^2/q^2$. By working inward from large $q$, the correlation function can be
used to determine inverse moments of the source function, $1/r$, $1/r^2$, etc.

\begin{figure}
\centerline{\includegraphics[width=0.6\textwidth]{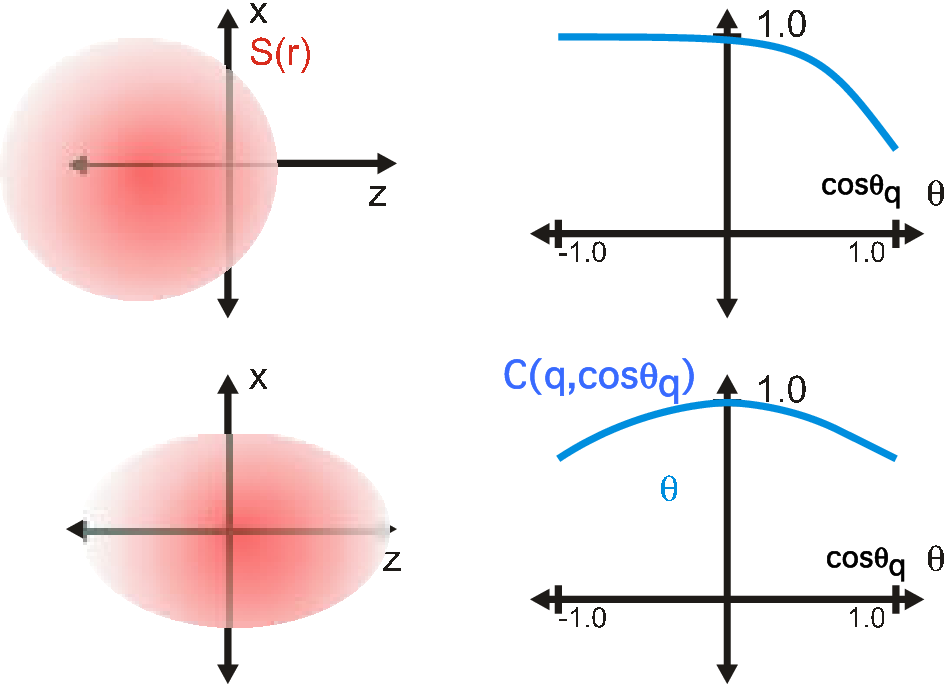}}
\caption{\label{fig:coulcartoon}
Coulomb repulsion suppresses trajectories where the outgoing relative momenta is
anti-parallel to the original separation. For sources biased to one side, as
illustrated in the upper panel, the correlation function is suppressed for
opposite directions. For sources with a quadrapole distortion along a given
axis, the distortion is strongest for emission parallel or anti-parallel to the
axis as illustrated in the lower panel.}
\end{figure}

The $\cos\theta_{qr}$ dependence in Eq. (\ref{eq:classcoul}) can be exploited
to extract shape information about the source. This sensitivity is driven by the
scattering of the Coulomb trajectories which lowers the population for
angles where ${\bf q}$ leaves anti-parallel to ${\bf r}$ as illustrated in
Fig. \ref{fig:coulcartoon}. 

\begin{figure}
\centerline{
\includegraphics[width=0.45\textwidth]{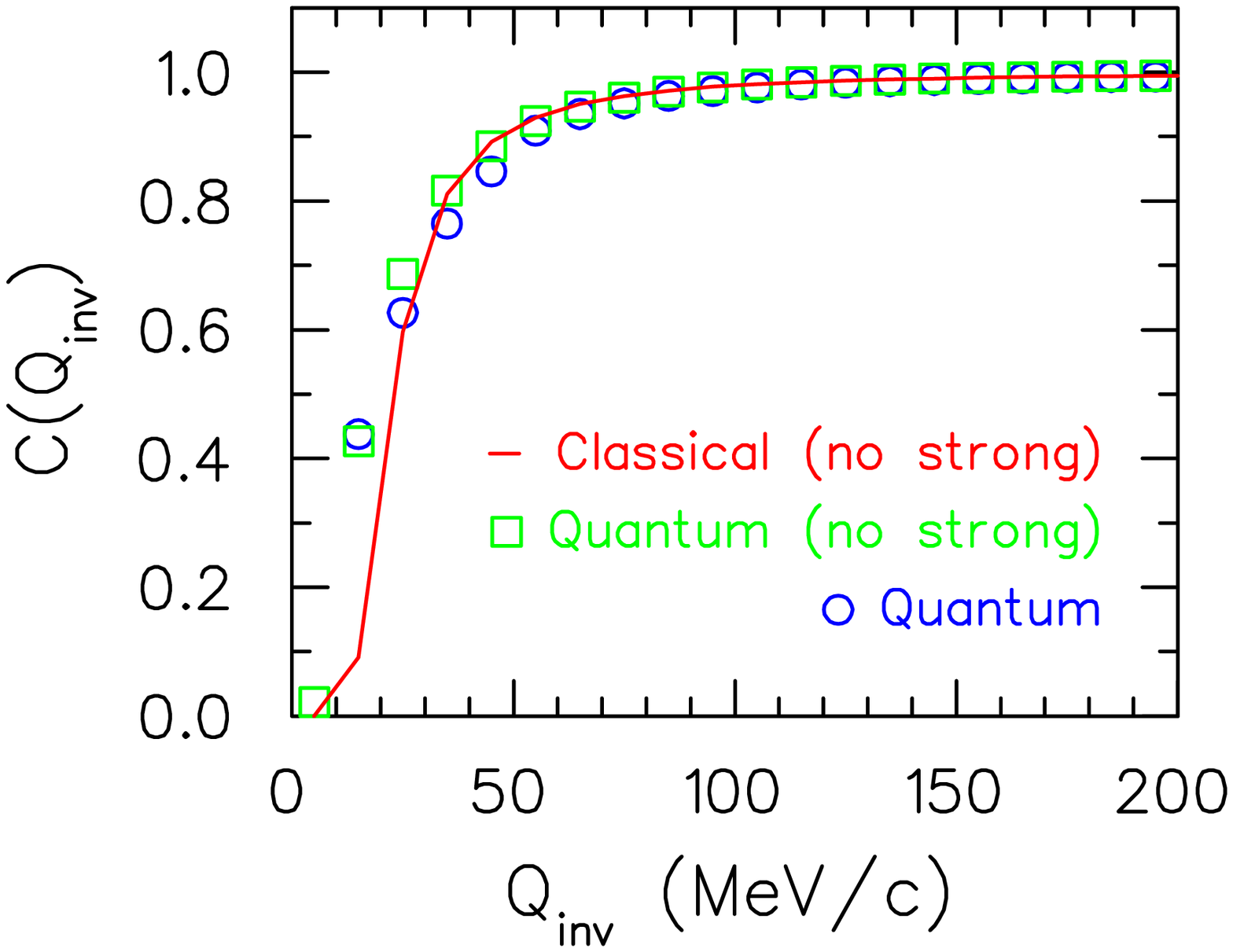}
\includegraphics[width=0.45\textwidth]{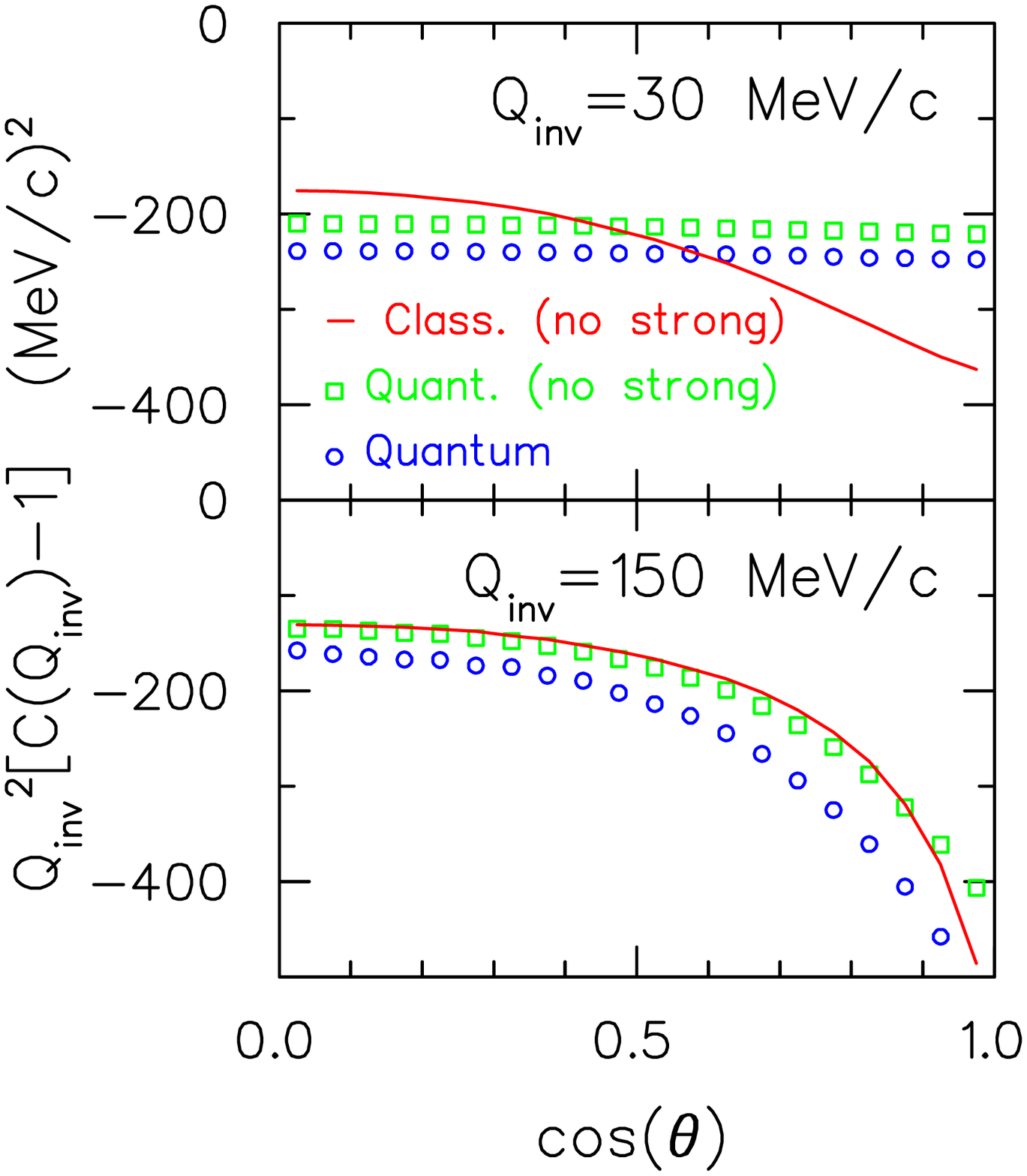}}
\caption{\label{fig:corrkk} Proton-kaon correlation functions are shown for a
  $4\times 4\times 8$ fm Gaussian source. Quantum effects mute the sensitivity
  predicted from classical Coulomb trajectories. This especially true at low
  $Q=2q$. In the right-side panel, the correlation functions are scaled by
  $Q^2$ to counter the $1/Q^2$ fall-off. Incorporating strong interactions into
  the quantum calculations also affects the correlation.  }
\end{figure}

The quantum Coulomb wave function is more complex, and adds a additional
sensitivity to $qr/\hbar$ in addition to $x$ and $\cos\theta_{qr}$. Quantum
effects tend to smear out some of the sensitivity for $qr/\hbar \sim 1$
\cite{petriconi}. This especially harmful for small reduced masses, such as
$\pi\pi$ correlations, since the correlations are smaller and need to be
analyzed at smaller $q$. In fact, for cases where the Bohr radius is much
larger than the source size ($a_0=1/\mu e^2=390$ fm for $\pi\pi$), the effect
of Coulomb is a simple Gamow factor which is independent of $r$ and thus
provides no information. Figure \ref{fig:corrkk} shows correlations $pK^+$
correlations for a $4\times 4\times 8$ fm Gaussian source. Compared to classical
correlations quantum correlations are somewhat muted, especially the shape
information which is determined by looking at the dependence on the direction
of ${\bf q}$. The muting is small for larger $q$, but there the correlation is
small. Nonetheless, correlations are strong enough near 100 $MeV/c$ to
determine shape characteristics of the source. The analogous calculations for
Fig. \ref{fig:corrkk} can be performed for any pair. Any interaction can be
exploited to determine shape. For instance $p\Lambda$ interactions, which have
no Coulomb, also provide sensitivity to shape through shadowing.

In Figure \ref{fig:corrkk} the anisotropy of the source was seen by plotting
the correlation function as a function of $\cos\theta_q$, where $\theta_q$ was
measured relative to the elongated axis. For more complicated three-dimensional
shapes, especially where the source may not have reflection symmetries,
determining characteristics of the shape can be more difficult. Spherical and
Cartesian harmonics \cite{danielewicz} can then be used to express the
correlation function in terms of a few coefficients at small $\ell$. These
expansion coefficients are themselves functions of $q$ and can be related to
expansion coefficients for the source functions.
\begin{equation}
\label{eq:ylmmaster}
R_{\ell,m}(q)=\int 4\pi r^2dr~
{\mathcal K}_\ell(q,r)~S_{\ell,m}(r),
\end{equation}
where these quantities are related to the correlation function, wave function
and source distribution by the following definitions,
\begin{eqnarray}
\label{eq:ylmmaster_defs}
R_{\ell,m}(q)&\equiv&
(4\pi)^{-1/2}\int d\Omega_q R({\bf q}) Y_{\ell,m}(\Omega_q),\\
\nonumber
S_{\ell,m}(r)&\equiv&
(4\pi)^{-1/2}\int d\Omega_r S({\bf r}) Y_{\ell,m}(\Omega_r),\\
\nonumber
{\mathcal K}_\ell(q,r)&\equiv&\frac{1}{2}\int d\cos\theta_{qr} \left[
|\phi(q,r,\cos\theta_{qr})|^2-1\right] P_\ell(\cos\theta_{qr}).
\end{eqnarray}
The convenience of this expression is that the expansion coefficients labeled
$\ell$ and $m$ are related on a one-to-one basis with the corresponding
coefficients in the source function.

The disadvantage with using spherical harmonics is that it can be difficult to
physically visualize the distortion associated with a given $\ell$ and $m$,
particularly for $\ell>2$. This can be remedied by using Cartesian
harmonics. Cartesian harmonics of order $\ell=\ell_x+\ell_y+\ell_z$ are linear
combinations of spherical harmonics of the same order. Each Cartesian harmonic
is of the form,
\begin{equation}
{\mathcal A}_{\vec\ell}(\hat{n})=n_x^{\ell_x}n_y^{\ell_y}n_z^{\ell_z}+{\rm
  ~lower~orders~of~}n_x,n_y,n_z.
\end{equation}
Some examples are given in Table \ref{table:cartesian}. The disadvantage of
Cartesian harmonics is that they are not orthonormal within a given $\ell$
multiplet. But, the $(2\ell+1)$ harmonics with $\ell_x=0,1$ can be used to
easily generate the other functions with the identity, ${\mathcal
  A}_{\ell_x,\ell_y,\ell_z}= -{\mathcal A}_{\ell_x-2,\ell_y+2,\ell_z}-{\mathcal
  A}_{\ell_x-2,\ell_y,\ell_z-2}$. Since Cartesian harmonics can be expressed as
a linear combination of spherical harmonics of the same $\ell$,
\begin{equation}
\label{eq:cartesianmaster}
R_{\vec{\ell}}(q)
=\int 4\pi r^2dr~{\mathcal K}_\ell(q,r)
~S_{\vec{\ell}}(r),
\end{equation}
where the coefficients are defined,
\begin{eqnarray}
\label{eq:cartesianmaster_defs}
R_{\vec{\ell}}(q)&\equiv&
\frac{(2\ell+1)!!}{\ell!}\int \frac{d\Omega_q}{4\pi}
R({\bf q}) {\mathcal A}_{\vec\ell}(\Omega_q),\\
S_{\vec{\ell}}(q)&\equiv&
\frac{(2\ell+1)!!}{\ell!}\int \frac{d\Omega_q}{4\pi}
S({\bf q}) {\mathcal A}_{\vec\ell}(\Omega_q).
\end{eqnarray}
References \cite{danielewicz,applequist} provide many properties of Cartesian
harmonics and gives expressions for orthonormality, for transforming to
spherical harmonics, and for generating moments.

\begin{table}
\caption{\label{table:cartesian} Cartesian harmonics for $\ell<=4$. Other
harmonics can be found by swapping indices on both sides of the
equation, e.g., $x\leftrightarrow y$. For example, given ${\mathcal
A}_{2,1,0}=n_x^2n_y-n_y/5$, swapping $y\leftrightarrow z$ gives ${\mathcal
A}_{2,0,1}=n_x^2n_z-n_z/5$.}
\begin{tabular}{|c|c|}\hline
${\mathcal A}_{1,0,0}=n_x$ 
& ${\mathcal A}_{1,1,1}=n_xn_yn_z$ \\
${\mathcal A}_{2,0,0}=n_x^2-(1/3)$ 
& $A_{4,0,0}=n_x^4-(6/7)n_x^2+(3/35)$ \\
${\mathcal A}_{1,1,0}=n_xn_y$ 
& ${\mathcal A}_{3,1,0}=n_x^3n_y-(3/7)n_xn_y$\\
$A_{3,0,0}=n_x^3-(3/5)n_x$ 
&  ${\mathcal A}_{2,2,0}=n_x^2n_y^2-(1/7)n_x^2-(1/7)n_y^2+(1/35)$\\
${\mathcal A}_{2,1,0}=n_x^2n_y-(1/5)n_y$ 
& ${\mathcal A}_{2,1,1}=n_x^2n_yn_z-(1/7)n_yn_z$\\ \hline
\end{tabular}
\end{table}

\begin{figure}
\centerline{\includegraphics[width=0.73\textwidth]{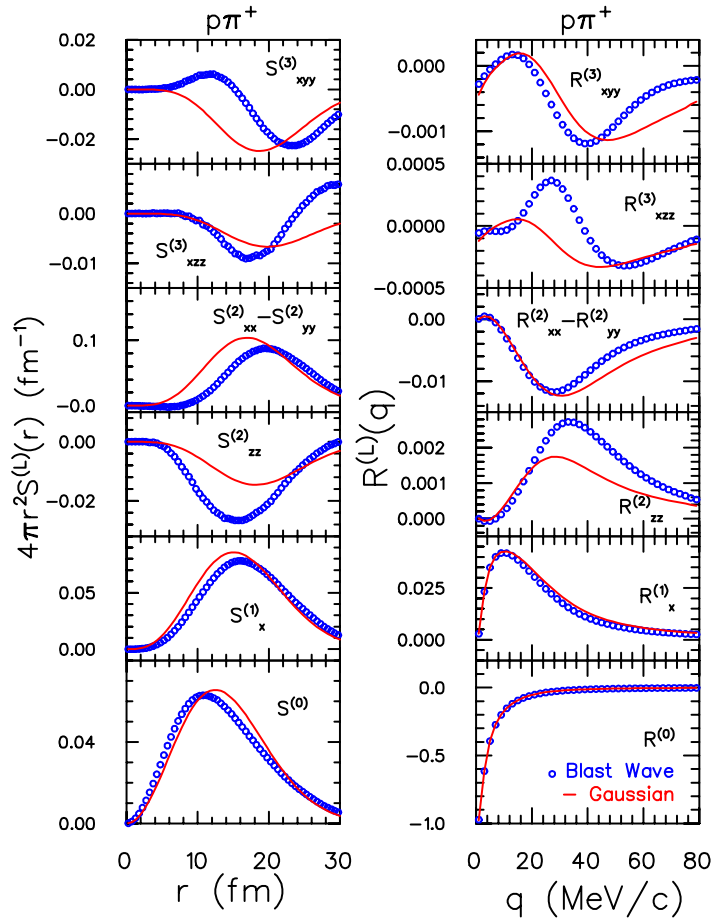}}
\caption{\label{fig:blast} Decompositions in Cartesian harmonics of $p\pi^+$
  source and correlation functions are shown for a blast-wave source (blue
  circles). A Gaussian source (Red), chosen to reproduce the integrated
  moments, does a poor job reproducing the source and correlation functions for
  some moments.  }
\end{figure}
Results for various moments were generated for $p\pi$ correlations from a
blast-wave model and are displayed in Fig. \ref{fig:blast}. The blast-wave
temperature was chosen to be 120 MeV, the radius was 13 fm and the collective
velocity was 0.7$c$. Since lighter particles tend to originate from deeper
inside the blast wave, the source functions have strong dipole moments, which
are manifest in the $\ell_x=1$ Cartesian projection, where $x$ refers to the
outward direction. The $\ell=2$ moments describe the elliptic anisotropies,
whereas the $\ell=1$ terms can be more directly identified with the offset of
the center of the ellipse. The $(\ell_x=1,\ell_y=0,\ell_z=2)$ term is driven
both by the combination of the offset and elliptic asymmetries and by the
boomerang shape caused by the inside-outside geometry of the cascade. The
non-Gaussian nature of the shape is seen by comparison with the best-fit
Gaussian source. Gaussians are especially bad for fitting the behavior of the
source function at larger $r$. Furthermore, the one-to-one correspondence for
labels of the correlations and source function makes source imaging, which has
been applied for angle-integrated correlations
\cite{browndanielewicz,panitkin,verde}, tenable for higher harmonics.

One of the great benefits of analyses based on spherical or Cartesian harmonics
is that the shape can be studied as a function of $r$. For instance, at RHIC
the source functions should have exponential tails along the beam axis due to
the boost-invariant nature of the expansion, and perhaps in the outward
direction due to resonances. The higher harmonics should then be pronounced at
large $r$ due to the decreasing justification of Gaussian fits. This is
apparent in Fig. \ref{fig:blast}

Analyses such as those illustrated here can be performed for nearly every
species measured at RHIC. For instance, with the last data run at RHIC, there
should be enough data to analyze $\Sigma\pi$ correlations. There are dozens of
possible analyses. In any model incorporating thermalization and collective
flow, source sizes and shapes for different species should follow simple
systematic trends driven by the particle masses. Observing such trends would be
of tremendous importance in solidifying our understanding of collision dynamics
at RHIC.

\section*{Acknowledgments} This work was supported by the U.S. Dept. of Energy, 
Grant  DE-FG02-03ER41259.

\end{document}